\documentclass[aps,preprint]{revtex4}%
\usepackage{amsfonts}
\usepackage{amsmath}
\usepackage{amssymb}
\usepackage{graphicx}%
\begin{document}
\preprint{For PA}
\title{Constructing non-equilibrium statistical ensemble formalism based on Subdynamics}
\author{Bi Qiao$^{a,b,c}$}
\affiliation{$^{a}$Physics Department, Science school of Wuhan University of Technology,
Wuhan 430070, China; $^{b}$Center of Advanced Nanotechnology, University of
Toronto, Toronto M5S 3E4, Canada; $^{c}$China Institute of Atomic Energy, P.O.
Box 275-27, Beijing 102413, China. }

\begin{abstract}
In this work, we present a general non-equilibrium ensemble formalism based on
the subdynamic equation (SKE). The constructing procedure is to use a
similarity transformation between Gibbsian ensemble formalism and the
non-equilibrium ensemble formalism. The obtained density distribution is a
projected one that can represent essence part of (irreversible) evolution of
the density distribution, by which a generalized reduced density distribution
for the quantum canonical ensembles is studied and applications in Cayley tree
and spin network are discussed.

\end{abstract}
\maketitle

\section{Introduction}

Since Gibbs synthesized a general equilibrium statistical ensemble theory,
many theorists have attempted to generalized the Gibbsian theory to
non-equilibrium phenomena domain, however the status of the theory of
non-equilibrium phenomena can not be said as firm as well established as the
Gibbian ensemble theory, although great works have done by numerous
authors$^{\text{\cite{q4}-\cite{q12}}}$. The number of references along this
line of research is too numerous to cite them all here, we just mention three
significant progresses: the relevant ensembles theory presented by Zubarev,
Morozov and R\"{o}pke$^{\text{\cite{0}}}$, the Jaynes' predictive statistical
mechanics approach$^{\text{\cite{v}}}$, and the generalized Gibbsian ensembles
theory based on the Boltzmann kinetic equation presented by Chan
Eu$^{\text{\cite{1}}}$. So far the obtained non-equilibrium statistical
density distribution formulas for the ensembles do not satisfy the original
Liouville equation. Some researchers for that reason believe that the
Liouville equation must have an extra term which satisfies a set of conditions
assuring its irreversibility and existence of conservation laws if the Gibbs
ensemble theory is generalized to the non-equilibrium phenomena domain based
on the Liouville equation. But how is it possible to find this extra term
which possesses universal irreversible characteristic to satisfy numerous
requirements from a large body of models? This means the efforts of many
school until now have not produced a universal ensemble theory for
non-equilibrium phenomena which is comparable to the Gibbian ensemble theory
for equilibrium phenomena.

In this work, we present a non-equilibrium statistical ensemble formalism
based on a subdynamic kinetic equation (SKE) rooted from the Brussels-Austin
school$^{\text{\cite{3}-\cite{4}}}$ and followed by some up-to-date
works$^{\text{\cite{5}-\cite{5b}}}$. The advantage of the scheme is that SKE
intertwines with the original Liouville equation by a similarity
transformation. If the similarity transformation is non-unitary, the SKE can
describe the irreversible process, otherwise, it describes the reversible
process as an equivalent equation of the original Liouville equation. Although
there exist\ several different approaches to construct SKE, that can be found
in some publications$^{\text{\cite{3}-\cite{5b}}}$, here considering reader
may be not familiar with formalism of subdynamics, we try to start from an update-introduction.

\section{Subdynamics Formalism}

Let a quantum system $S$ be coupled (may be strongly) to a thermal reservoir
$B$, $H_{S}\left(  t\right)  $, $H_{B}$, and $H_{int}$ denote the Hamiltonian
of the system $S$, the Hamiltonian of the thermal reservoir $B$, and the
interaction between $S$ and $B$, respectively. The total Hamiltonian $H\left(
t\right)  $ of the system plus the reservoir can be expressed as $H_{S}\left(
t\right)  \otimes I_{B}$ + $I_{S}\otimes H_{B}$ + $H_{int}$, and the
corresponding quantum Schr\"{o}dinger equation and Liouville equation are
\begin{equation}
i\frac{\partial f}{\partial t}=H\left(  t\right)  f,\label{1}%
\end{equation}
and
\begin{equation}
i\frac{\partial\rho}{\partial t}=\left[  H\left(  t\right)  ,\rho\right]
,\label{2}%
\end{equation}
where $\rho=\left\vert f\right\rangle \left\langle f\right\vert $ is a density
operator for the total system. Then one can introduce an orthonormal projector
$P_{kj}$ with $Q_{kj}=1-P_{kj}$ so that%
\begin{align}
P_{kj}H\left(  t\right)  P_{kj} &  =P_{kj}H_{0}\left(  t\right)
P_{kj},\label{3}\\
P_{kj}H\left(  t\right)  Q_{kj} &  =P_{kj}H_{int}Q_{kj}.\label{4}%
\end{align}
Then the total Hamiltonian $H\left(  t\right)  =H_{S}\left(  t\right)
+H_{B}+H_{in}$ can be expressed as%
\begin{equation}
H\left(  t\right)  =PH\left(  t\right)  P+PH\left(  t\right)  Q+QH\left(
t\right)  P+QH\left(  t\right)  Q,\label{7}%
\end{equation}
and a corresponding projected matrix is represented as%
\begin{equation}
\left(
\begin{array}
[c]{cc}%
P_{kj}H\left(  t\right)  P_{kj} & Q_{kj}H\left(  t\right)  P_{kj}\\
P_{kj}H\left(  t\right)  Q_{kj} & Q_{kj}H\left(  t\right)  Q_{kj}%
\end{array}
\right)  .\label{8}%
\end{equation}
The eigenvalue problem can be written as%
\begin{equation}
\left(
\begin{array}
[c]{cc}%
P_{kj}H\left(  t\right)  P_{kj} & Q_{kj}H\left(  t\right)  P_{kj}\\
P_{kj}H\left(  t\right)  Q_{kj} & Q_{kj}H\left(  t\right)  Q_{kj}%
\end{array}
\right)  \left(
\begin{array}
[c]{c}%
\phi_{kj}\\
\phi_{k^{\prime}j^{\prime}}%
\end{array}
\right)  =\left(
\begin{array}
[c]{cc}%
E_{kj}\left(  t\right)   & 0\\
0 & E_{k^{\prime}j^{\prime}}\left(  t\right)
\end{array}
\right)  \left(
\begin{array}
[c]{c}%
\phi_{kj}\\
\varphi_{k^{\prime}j^{\prime}}%
\end{array}
\right)  ,\label{9}%
\end{equation}
giving%
\begin{align}
P_{kj}H\left(  t\right)  P_{kj}\phi_{kj}+P_{kj}H\left(  t\right)  Q_{kj}%
\phi_{k^{\prime}j^{\prime}} &  =E_{kj}\left(  t\right)  \phi_{kj}%
,\label{b13}\\
Q_{kj}H\left(  t\right)  P_{kj}\phi_{kj}+Q_{kj}H\left(  t\right)  Q_{kj}%
\phi_{k^{\prime}j^{\prime}} &  =E_{k^{\prime}j^{\prime}}\left(  t\right)
\phi_{k^{\prime}j^{\prime}}.\label{b15}%
\end{align}
From Eqs.(\ref{b15}) and (\ref{b13}), one can solve $\phi_{k^{\prime}%
j^{\prime}}$ and $\phi_{kj}$ respectively as
\begin{align}
\phi_{k^{\prime}j^{\prime}} &  =\left(  E_{k^{\prime}j^{\prime}}\left(
t\right)  -Q_{kj}H\left(  t\right)  Q_{kj}\right)  ^{-1}Q_{kj}H\left(
t\right)  P_{kj}\phi_{kj}\label{b16}\\
&  =C_{kj}\phi_{kj},\nonumber
\end{align}%
\begin{align}
\phi_{kj} &  =\left(  E_{kj}\left(  t\right)  -P_{kj}H\left(  t\right)
P_{kj}\right)  ^{-1}P_{kj}H\left(  t\right)  Q_{kj}\phi_{k^{\prime}j^{\prime}%
}\label{b16b}\\
&  =C_{k^{\prime}j^{\prime}}\phi_{k^{\prime}j^{\prime}}.\nonumber
\end{align}
Substituting Eq.(\ref{b16}) into Eq.(\ref{b13}) and Eq.(\ref{b16b}) into
Eq.(\ref{b15}) respectively gives%
\begin{align}
\left(  P_{kj}H\left(  t\right)  P_{kj}+P_{kj}H\left(  t\right)  C_{k^{\prime
}j^{\prime}}\left(  t\right)  \right)  \phi_{kj} &  =\Theta_{k^{\prime
}j^{\prime}}\left(  t\right)  \phi_{kj}=E_{kj}\phi_{kj},\label{17}\\
\left(  Q_{kj}H\left(  t\right)  P_{kj}+Q_{kj}H\left(  t\right)  C_{kj}\left(
t\right)  \right)  \phi_{k^{\prime}j^{\prime}} &  =\Theta_{kj}\left(
t\right)  \phi_{k^{\prime}j^{\prime}}=E_{k^{\prime}j^{\prime}}\left(
t\right)  \phi_{k^{\prime}j^{\prime}},\label{17b}%
\end{align}
where introducing a creation (destruction) correlation operator (as a type of
resolvent) as%
\begin{align}
C_{kj}\left(  t\right)   &  =Q_{kj}C_{kj}\left(  t\right)  P_{kj}=\left(
E_{k^{\prime}j^{\prime}}\left(  t\right)  -Q_{kj}H\left(  t\right)
Q_{kj}\right)  ^{-1}Q_{kj}H\left(  t\right)  P_{kj},\label{18a}\\
C_{k^{\prime}j^{\prime}}\left(  t\right)   &  =P_{kj}C_{k^{\prime}j^{\prime}%
}\left(  t\right)  Q_{kj}=\left(  E_{kj}\left(  t\right)  -P_{kj}H\left(
t\right)  P_{kj}\right)  ^{-1}P_{kj}H\left(  t\right)  Q_{kj}.\label{18b}%
\end{align}
This shows that the $\left(  \phi_{kj},\phi_{k^{\prime}j^{\prime}}\right)  $
are the eigenvectors of the $\Theta_{k^{\prime}j^{\prime}}\left(  t\right)  $
and $\left(  E_{kj}\left(  t\right)  ,E_{k^{\prime}j^{\prime}}\left(
t\right)  \right)  $ are the eigenvalues of $\Theta_{k^{\prime}j^{\prime}%
}\left(  t\right)  $ and $H\left(  t\right)  $. This allows one to presume
that the eigenvector of $H\left(  t\right)  $ is given by $f_{kj}$ with the
same eigenvalue $E_{kj}\left(  t\right)  $,
\begin{equation}
H\left(  t\right)  \left(
\begin{array}
[c]{c}%
\left\vert f_{kj}\right\rangle \\
\left\vert f_{k^{\prime}j^{\prime}}\right\rangle
\end{array}
\right)  =\left(
\begin{array}
[c]{c}%
Z_{kj}\left(  t\right)  \left\vert f_{kj}\right\rangle \\
Z_{k^{\prime}j^{\prime}}\left(  t\right)  \left\vert f_{k^{\prime}j^{\prime}%
}\right\rangle
\end{array}
\right)  ,\label{19}%
\end{equation}
then one can find by using Eqs.(\ref{9}-\ref{17b}),%
\begin{align}
\left(
\begin{array}
[c]{c}%
P_{kj}H\left(  P_{kj}+Q_{kj}\right)  f_{kj}\\
Q_{jk}H\left(  P_{kj}+Q_{kj}\right)  f_{kj}%
\end{array}
\right)   &  =\left(
\begin{array}
[c]{cc}%
P_{kj}HP_{kj} & P_{kj}HQ_{kj}\\
Q_{kj}HP_{kj} & Q_{kj}HQ_{kj}%
\end{array}
\right)  \left(
\begin{array}
[c]{c}%
P_{kj}f_{kj}\\
Q_{kj}f_{kj}%
\end{array}
\right)  \label{aa1}\\
&  =\left(
\begin{array}
[c]{cc}%
E_{kj}\left(  t\right)   & 0\\
0 & E_{k^{\prime}j^{\prime}}\left(  t\right)
\end{array}
\right)  \left(
\begin{array}
[c]{c}%
\phi_{kj}\\
\phi_{k^{\prime}j^{\prime}}%
\end{array}
\right)  \nonumber\\
&  =\left(
\begin{array}
[c]{cc}%
\Theta_{k^{\prime}j^{\prime}} & 0\\
0 & \Theta_{kj}%
\end{array}
\right)  \left(
\begin{array}
[c]{c}%
\phi_{kj}\\
\phi_{k^{\prime}j^{\prime}}%
\end{array}
\right)  .\nonumber
\end{align}
This gives some interesting relations as%
\begin{align}
P_{kj}f_{kj} &  =\phi_{kj}=C_{kj}\phi_{k^{\prime}j^{\prime}}=C_{kj}%
Q_{kj}f_{kj},\label{d1}\\
Q_{kj}f_{kj} &  =\phi_{k^{\prime}j^{\prime}}=C_{k^{\prime}j^{\prime}}\phi
_{kj}=C_{k^{\prime}j^{\prime}}P_{kj}f_{kj},\label{d2}%
\end{align}
and%
\begin{equation}
f_{kj}=P_{kj}\phi_{kj}+Q_{kj}\phi_{k^{\prime}j^{\prime}}=\left(
P_{kj}+C_{k^{\prime}j^{\prime}}\right)  \phi_{kj}.\label{d3}%
\end{equation}
Using Eq.(\ref{d2}), by introducing $\Pi_{kj}\left(  t\right)  =\left\vert
f_{kj}\left(  t\right)  \right\rangle \left\langle f_{kj}\left(  t\right)
\right\vert $ as an eigen-projector of $H\left(  t\right)  $, one can
construct a Schr\"{o}dinger type of kinetic equation for each projected state
$P_{kj}\Pi_{kj}\left(  t\right)  \left\vert \phi\left(  t\right)
\right\rangle $ as%
\begin{align}
i\frac{\partial}{\partial t}P_{kj}\Pi_{kj}\left(  t\right)  \left\vert
\phi\left(  t\right)  \right\rangle  &  =iP_{kj}\left[  \left(  \frac
{\partial}{\partial t}\Pi_{kj}\left(  t\right)  \right)  \left\vert
\phi\left(  t\right)  \right\rangle +\Pi_{kj}\left(  t\right)  \frac{\partial
}{\partial t}\left\vert \phi\left(  t\right)  \right\rangle \right]
\label{bb6}\\
&  =P_{kj}\left\{  \left[  H\left(  t\right)  ,\Pi_{kj}\left(  t\right)
\right]  \left\vert \phi\left(  t\right)  \right\rangle +\Pi_{kj}\left(
t\right)  H\left(  t\right)  \left\vert \phi\left(  t\right)  \right\rangle
\right\}  \nonumber\\
&  =P_{kj}H\left(  t\right)  \left(  P_{kj}+Q_{kj}\right)  \Pi_{kj}\left(
t\right)  \left\vert \phi\left(  t\right)  \right\rangle \nonumber\\
&  =P_{kj}H\left(  t\right)  \left(  P_{kj}+C_{kj}\left(  t\right)  \right)
\Pi_{kj}\left(  t\right)  \left\vert \phi\left(  t\right)  \right\rangle
,\nonumber
\end{align}
where, for more generality, $\Pi_{kj}\left(  t\right)  $ can be understood as
$\left\vert f_{kj}\left(  t\right)  \right\rangle \left\langle \widetilde
{f}_{kj}\left(  t\right)  \right\vert $ in which $f_{kj}\left(  t\right)
\in\Phi$ (dense subspace) and $\widetilde{f}_{kj}\left(  t\right)  \in
\Phi^{\times}$ (generalized dual subspace of $\Phi$) are defined in a Rigged
Hilbert space, $\Phi\subset\mathcal{H}\subset\Phi^{\times}$. This can
generally provide a Schr\"{o}dinger type of subdynamics kinetic equation
(SSKE) expressed as%
\begin{align}
i\frac{\partial}{\partial t}\phi_{proj}\left(  t\right)   &  =\Theta\left(
t\right)  \phi_{proj}\left(  t\right)  ,\label{21}\\
-i\frac{\partial}{\partial t}\widetilde{\phi}_{proj}\left(  t\right)   &
=\Theta\left(  t\right)  \widetilde{\phi}_{proj}\left(  t\right)  ,\label{21b}%
\end{align}
with%
\begin{equation}
\Theta\left(  t\right)  =\sum_{kj}P_{kj}H\left(  t\right)  \left(
P_{kj}+C_{kj}\left(  t\right)  \right)  ,\label{22}%
\end{equation}
where $\phi_{proj}$\ and $\widetilde{\phi}_{proj}\left(  t\right)  $ are
defined as
\begin{align}
\left\vert \phi_{proj}\left(  t\right)  \right\rangle  &  =\sum_{kj}P_{kj}%
\Pi_{kj}\left(  t\right)  \left\vert \phi\left(  t\right)  \right\rangle
,\label{bq2}\\
\left\langle \widetilde{\phi}_{proj}\left(  t\right)  \right\vert  &
=\sum_{kj}\left\langle \widetilde{\phi}\left(  t\right)  \right\vert P_{kj}%
\Pi_{kj}\left(  t\right)  ,\label{bq2b}%
\end{align}
and $\phi\left(  t\right)  $ or $\widetilde{\phi}\left(  t\right)  $ is a
solution of the original Schr\"{o}dinger equation in the Rigged Hilbert space.
Furthermore, by replacing $\rho_{proj}\left(  t\right)  =\left\vert
\phi_{proj}\left(  t\right)  \right\rangle \left\langle \widetilde{\phi
}_{proj}\left(  t\right)  \right\vert $, and using the above SSKE, a
Liouvillian type of kinetic equation (LSKE) can be derived by%
\begin{align}
i\frac{\partial}{\partial t}\rho_{proj}\left(  t\right)   &  =\left(
i\frac{\partial}{\partial t}\left\vert \phi_{proj}\left(  t\right)
\right\rangle \right)  \left\langle \phi_{proj}\left(  t\right)  \right\vert
+\left\vert \phi_{proj}\left(  t\right)  \right\rangle \left(  i\frac
{\partial}{\partial t}\left\langle \phi_{proj}\left(  t\right)  \right\vert
\right)  \label{eqn15}\\
&  =\Theta\left(  t\right)  \left\vert \phi_{proj}\left(  t\right)
\right\rangle \left\langle \phi_{proj}\left(  t\right)  \right\vert
-\left\vert \phi_{proj}\left(  t\right)  \right\rangle \left\langle
\phi_{proj}\left(  t\right)  \right\vert \Theta\left(  t\right)  \nonumber\\
&  =\left[  \Theta\left(  t\right)  ,\rho_{proj}\left(  t\right)  \right]
.\nonumber
\end{align}

The construction of SSKE or LSKE in subspace can be related to the original
Schr\"{o}dinger or Liouville equation$^{\text{\cite{3},\cite{5}}}$. For
instance, using the relation (\ref{d3}) one have the spectral representation
of $H\left(  t\right)  $ related to $\Theta\left(  t\right)  $ as%
\begin{align}
H\left(  t\right)   &  =\sum_{kj}Z_{kj}\left(  t\right)  \left\vert
f_{kj}\right\rangle \left\langle \widetilde{f}_{kj}\right\vert \label{16}\\
&  =\sum_{kj}Z_{kj}\left(  t\right)  \left(  P_{kj}+C_{kj}\left(  t\right)
\right)  \left\vert \phi_{kj}\right\rangle \left\langle \phi_{kj}\right\vert
\left(  P_{kj}+D_{kj}\left(  t\right)  \right) \nonumber\\
&  =\Omega\left(  t\right)  \Theta\left(  t\right)  \Omega^{-1}\left(
t\right)  ,\nonumber
\end{align}
where $D_{kj}\left(  t\right)  =C_{kj}^{\dagger}\left(  t\right)  $, and
$\Omega\left(  t\right)  =\sum_{kj}\left(  P_{kj}+C_{kj}\left(  t\right)
\right)  $. The creation operator, $C_{\nu}=\frac{1}{Z-Q_{\nu}HQ_{\nu}}Q_{\nu
}HP_{\nu}=\left(  D_{\nu}\right)  ^{\dagger}$, creates the $Q_{\nu}$-part of
$\Pi_{\nu}$ from the $P_{\nu}$-part. While $\Theta=H_{0}+\lambda H_{1}C$ is
called collision operator$^{\text{\cite{77}}}$. The physical meaning\ of
$\rho_{proj}$ is that it represents the "vacuum" part of the "dynamic" part of
the original density operator $\rho$, which describes the essence of
(irreversible) evolution of the density $\rho$ in its own
subspace$^{\text{\cite{7}}}$. The second order approximation of $\Theta$ with
respect to $\lambda$ corresponds to the Master equation$^{\text{\cite{6}}}$.
Moreover, the Boltzmann, Pauli, and Fokker-Planck equations of kinetic theory
and Brownian motion can also be derived by using some approximation of
$\Theta$ $^{\text{\cite{7}}}$. The creation operator and destruction operator
can also be calculated by using operator algebra to perform several different
approaches. One of interested approaches to obtain the recurrent formulas is%

\begin{align}
C_{\nu}  &  =i\int_{0}^{\pm\infty}d\tau U\left(  \tau\right)  \lambda\left(
C_{\nu}-Q_{\nu}\right)  H\left(  P_{\nu}+C_{\nu}\right)  U\left(
-\tau\right)  ,\label{23}\\
D_{\nu}  &  =i\int_{0}^{\pm\infty}d\tau U\left(  \tau\right)  \lambda\left(
P_{\nu}+D_{\nu}\right)  H\left(  P_{\mu}-D_{\nu}P_{\mu}\right)  U\left(
-\tau\right)  , \label{24}%
\end{align}
where defining $U\left(  \tau\right)  =\exp\left(  -\tau H_{0}\right)  $.

\section{ Non-equilibrium statistical ensembles}

A marvelous remark is that the SKE seems to have the general property to
approach various kinetic equations or Master equations, which is beyond the
original Liouville equation. As previous mentioned, the Brussels-Austin group
have developed many important works for SKE in last two decades and have found
that SKE can intertwine with the original Liouville equation by a similarity
operator. If the similarity operator is unitary, the SKE is reversible, as an
equivalent representation of Liouville equation; if the similarity operator is
not unitary, the SKE is irreversible and the corresponding evolution is not
time symmetric. This means that the SKE can be as an appropriate kinetic
equation to describe the irreversible process, in which the evolution operator
is non-unitary on generalized functional space which is beyond the traditional
Hilbert (or Liouville) space. This motivates one to propose using the SKE to
construct a non-equilibrium statistical ensemble theory. The constructing
procedure may be quite simple by using the "similarity transformation
corresponding" between Gibbsian ensembles formalism based on the Liouville
equation and the non-equilibrium ensembles formalism based on SKE: if the
Hamiltonian corresponding to an expectation value, then the corresponding
expectation of the $\Theta$ operator should be
\begin{equation}
Tr\left(  H\rho\right)  =\left\langle H\right\rangle \longrightarrow Tr\left(
\Theta P\Pi\rho\right)  =\left\langle \Theta\right\rangle ,\label{25}%
\end{equation}
thus the related entropy tends to extremum, this allows one to present (by
extension) a new canonical ensemble distribution $\rho\left(  \theta
_{k}\right)  $ which is "vacuum" of "dynamic part" of the original
$\rho\left(  E_{k}\right)  $, as expressed by Balescu's book \cite{7},%

\begin{equation}
\rho\left(  E_{k}\right)  =Z^{-1}\left(  \beta,V,N\right)  \exp\left(  -\beta
E_{k}\right)  \longrightarrow\rho\left(  \theta_{k}\right)  =Z^{-1}\left(
\beta_{proj},V_{proj},N_{proj}\right)  \exp\left(  -\beta_{proj}\theta
_{k}\right)  \label{b2}%
\end{equation}
with the partition functions as%
\begin{equation}
Z\left(  \beta,V,N\right)  =\sum_{k}\exp\left(  -\beta E_{k}\right)
\longrightarrow Z\left(  \beta_{proj},V_{proj},N_{proj}\right)  =\sum_{k}%
\exp\left(  -\beta_{proj}\theta_{k}\right)  , \label{b3}%
\end{equation}%
\begin{equation}
\beta=\left(  k_{B}T\right)  ^{-1}\longrightarrow\beta_{proj}=\left(
k_{B}T_{proj}\right)  ^{-1}, \label{3a}%
\end{equation}
where $\theta_{k}$ is an eigenvalue of $\Theta$, $\beta_{proj}$ is extended as
function of position and time. In fact, suppose the density distribution in
quantum canonical system is given by%
\begin{equation}
\rho\left(  E_{kj}\right)  =\left\{  \frac{\exp\left(  -\beta E_{kj}\right)
}{\sum_{kj}\exp\left(  -\beta E_{kj}\right)  }\right.  , \label{26}%
\end{equation}
which gives the density operator $\rho$ as%
\begin{align}
\rho &  =\sum_{kj}\left\vert f_{kj}\right\rangle \frac{\exp\left(  -\beta
E_{kj}\right)  }{\sum_{kj}\exp\left(  -\beta E_{kj}\right)  }\left\langle
f_{kj}\right\vert \label{27}\\
&  =\frac{1}{Tr\exp\left(  -\beta H\right)  }\exp\left(  -\beta H\right)
\sum_{kj}\left\vert f_{kj}\right\rangle \left\langle f_{kj}\right\vert
\nonumber\\
&  =\frac{\exp\left(  -\beta H\right)  }{Tr\exp\left(  -\beta H\right)
}.\nonumber
\end{align}
Thus using the similarity transformation $\Omega$ one can obtain a projected
density operator $\rho_{proj}$ as%
\begin{align}
\rho_{proj}  &  =\Omega^{-1}\rho\Omega=\Omega^{-1}\frac{\exp\left(  -\beta
H\right)  }{Tr\exp\left(  -\beta H\right)  }\Omega\label{28}\\
&  =\frac{\exp\left(  -\Omega^{-1}\beta\Omega\Omega^{-1}H\Omega\right)
}{Tr\exp\left(  -\Omega^{-1}\beta\Omega\Omega^{-1}H\Omega\right)  }\nonumber\\
&  =\frac{\exp\left(  -\beta_{proj}\Theta\right)  }{Tr\exp\left(
-\beta_{proj}\Theta\right)  }.\nonumber
\end{align}
This gives a precise formula of the quantum canonical ensemble for a projected
density operator $\rho_{proj}$, which can be considered as generalizing the
equilibrium quantum canonical ensembles formula to the
non-equilibrium\ quantum canonical ensembles formula in the sense as (1) if
the similarity operator is unitary, then the new formula is just an effective
representation of the old equilibrium quantum canonical ensembles formula
because $\Theta$ or $H$ has the same spectral structure, (2) if the similarity
operator is non-unitary, then the new formula is an extension of the old
formula and the spectrum of $\Theta$ may appear to have complex spectral
structure that is impossible to get from the original self-adjoint operator
$H$ in the Hilbert space, which represents kind of non-equilibrium quantum
canonical ensembles formula and reflects irreversibility of the system, and
(3) if the similarity operator can be deduced by some approximations, such as
Markovian/non-markovian approximations, then the new formula can expose some
non-equilibrium characteristics, which can not be gained from the equilibrium
quantum ensemble formulas.

Thus it is obvious that the preceding constructed quantum formalism for
density operator $\rho\left(  \theta_{k}\right)  $ can be extended to the
classical statistical canonical ensemble by%
\begin{equation}
\rho\left(  \theta_{k}\right)  =Z^{-1}\left(  \beta_{proj},V_{proj}%
,N_{proj}\right)  \exp\left(  -\beta_{proj}\theta_{k}\right)  \label{r13}%
\end{equation}
with%
\begin{equation}
Z\left(  \beta_{proj},V_{proj},N_{proj}\right)  =\int\exp\left(  -\beta
_{proj}\Theta\right)  dx.\label{r14}%
\end{equation}
In the same way, the non-equilibrium grand canonical ensembles distribution
can also be constructed by%
\begin{equation}
\rho\left(  \theta_{k}\right)  =Z^{-1}\left(  \beta_{proj},\mu_{proj}%
,V_{proj}\right)  \exp\left[  -\beta_{proj}\theta_{k}-\mu_{proj}%
N_{proj}\right]  ,\label{r15}%
\end{equation}
where the partition function is given by
\begin{equation}
Z\left(  \beta_{proj},\mu_{proj},V_{proj}\right)  =\sum_{k}\int\exp\left[
-\beta_{proj}\theta_{k}-\mu_{proj}N_{proj}\right]  .\label{r16}%
\end{equation}
Furthermore, the general canonical ensembles distribution may be written by
\begin{equation}
\rho=Z^{-1}\exp\left[  -\beta_{proj}\theta_{k}-\mu_{proj}N_{proj}-\sum
_{k}\gamma_{k}\Gamma_{k}\right]  ,\label{r17}%
\end{equation}
where the thermodynamic meanings of the parameters $\gamma_{k}$, $\Gamma_{k}$
can be fixed by thermodynamic correspondence. Again, the physical meaning of
$\Theta$, $\beta_{proj}$, $\mu_{proj}$ and $N_{proj}=\Omega^{-1}N_{k}\Omega$
are also the "vacuum" of "dynamic part" of the corresponding parameters, which
can be functional of variable of the coordinate of the system; when the $k$
system in the ensemble tends to equilibrium, they tend to equilibrium $H_{k}$
$\beta$, $\mu$ and $N_{k}$, respectively. We want to emphasize again that in
the book of Balescue$^{\text{\cite{7}}}$ the "dynamic part" means essence part
of (irreversible) evolution of the density distribution, and the "vacuum"
means without correlations. His work and Brussels-Austin school late works
seem to show that the $\rho_{proj}$ plays an important or influential role in
the (irreversible) evolution of the system by extending it to the Rigged
Hilbert space or Rigged Liouville space$^{\text{\cite{9}}}$. Using this way
can one build a corresponding relation between equilibrium statistical
ensemble formalism and non-equilibrium statistical ensemble formalism? The
answer is confirmed because the original Hamiltonian of the system has
corresponding relation to the collision operator by the similarity
transformation. Thus the dynamic variables $Y$ are usually obtained by
calculated over the non-equilibrium statistical distribution $\rho\left(
\theta_{k}\right)  $ which is given by the proposed non-equilibrium
statistical ensemble formulas (\ref{b2}) or (\ref{r15}) or solution of the SKE
(\ref{eqn15}), $\left\langle Y\right\rangle =Tr\left(  Y\rho\left(  \theta
_{k}\right)  \right)  $. If the second order approximation of $\Theta$
corresponds to the Master equation, the Boltzmann equation, the Pauli
equation, or the Fokker-Planck equation, then $Tr\left(  Y\rho\left(
\theta_{k}\right)  \right)  $ should deliver the expectation of corresponding
physical value in the non-equilibrium ensembles. The Eq. (\ref{28}) can be as
starting base to get non-equilibrium statistical ensembles formulations for
irreversibility, as demonstration of application below.

\section{Applications}

The $\rho_{proj}$ can be a generalized reduced density operator by choosing an
appropriate projector $P$ defined as%
\begin{equation}
\frac{\exp\left(  -\beta H_{B}\right)  }{Tr_{B}\exp\left(  -\beta
H_{B}\right)  }Tr_{B}\rho=P\rho P,\label{29}%
\end{equation}
with
\begin{equation}
PH_{0}P=\frac{\exp\left(  -\beta H_{B}\right)  }{Tr_{B}\exp\left(  -\beta
H_{B}\right)  }Tr_{B}H_{0},\label{o3}%
\end{equation}
and $P+Q=1$. Then the relations can be proved by assuming $H_{0}$ is
diagonalized and $H_{int}$ is off-diagonalized:%
\begin{align}
PHP &  =PH_{0}P,PH_{int}Q=0,\label{o4}\\
PHQ &  =PH_{int}Q,PH_{0}Q=0.\label{o5}%
\end{align}
Thus, using Eqs.(\cite{d1}) and (\cite{d2}), one can introduce%
\begin{align}
Q\Pi &  =CP\Pi,\label{o6}\\
\Pi Q &  =\Pi PD.\label{o7}%
\end{align}
This gives%
\begin{equation}
\rho_{proj}=\left\vert P\Pi\phi\right\rangle \left\langle P\Pi\widetilde{\phi
}\right\vert ,\label{o8}%
\end{equation}
which is just a kind of generalization of the reduced density operator for the
open system. This means that a generalized Markovian (or non-markovian)
equation for the generalized reduced density operator, $\rho_{proj}$, in
quantum canonical ensembles can be given as the formula (\ref{28}). For
example a generalized Markovian equation may be derived by introducing
$z^{0}-QH_{0}Q$ to replace $z-QHQ$ in the creation operator $C$ to cancel some
memory effects of $C$,
\begin{equation}
\rho_{proj}=\frac{\exp\left(  -\beta_{proj}\Theta\right)  }{Tr\exp\left(
-\beta_{proj}\Theta\right)  },\label{bb2}%
\end{equation}
with%
\begin{equation}
\Theta=PHP+PHQ\frac{1}{z^{0}-QH_{0}Q}QHP,\label{bb1}%
\end{equation}
where $z^{0}$ is an eigenvalue of free Hamiltonian $H_{0}$. Furthermore, a
Markovian equation for the reduced density can be obtained by the second
approximation with respect to the coupling number $\lambda$ from the above
equation,
\begin{align}
\rho_{proj} &  =\left(  P+D\right)  \rho\left(  P+C\right)  \label{q1}\\
&  =\frac{\exp\left(  -\beta_{proj}\Theta\right)  }{Tr\exp\left(
-\beta_{proj}\Theta\right)  }\nonumber\\
&  \approx P\rho P+D\rho P+P\rho C+O\left(  \lambda^{4}\right)  \nonumber\\
&  \approx P\rho P+O\left(  \lambda^{2}\right)  \nonumber\\
&  =\frac{\exp\left(  -\beta H_{B}\right)  }{Tr_{B}\exp\left(  -\beta
H_{B}\right)  }Tr_{B}\rho\nonumber\\
&  =\frac{\exp\left(  -\beta_{proj}\Theta^{\prime}\right)  }{Tr\exp\left(
-\beta_{proj}\Theta^{\prime}\right)  },\nonumber
\end{align}
where
\begin{equation}
\Theta^{\prime}=PHP+PHC,\label{o11}%
\end{equation}%
\begin{align}
C &  =-i\lambda\int_{0}^{\pm\infty}d\tau U\left(  \tau\right)  QHPU\left(
-\tau\right)  \label{o12}\\
&  =-i\lambda\sum_{n}\frac{1}{z_{n}^{0}-QH_{0}Q}QHP,\nonumber
\end{align}
and
\begin{equation}
\beta_{proj}=P\beta P+O\left(  \lambda^{2}\right)  .\label{o13}%
\end{equation}

As an application of the above formalism to the irreversibility, let us
consider a Cayley tree system (immersed in a nose environment) subject to a
strong interaction from environment or an external field. The Cayley tree is a
loop-free network in which there exist three classes of nodes, they are (1)
the root nodes, which is at origin of the tree and has connectivity $m$, (2)
the nodes at interface with connectivity $1$, and (3) the nodes below the
interface with connectivity $m+1$. Suppose that the network start from the
root of the tree with nodes $i=1$, and link it to $m$ new nodes $i=2,3,\cdots
$, $m+1$, one can indicates each node with a subsequent number, $t_{i}$
indicating the time in which it arrives in the interface. At each time step,
one can choose a node to grow, which gives rise to $m$ new nodes.
Consequently, the interface of the tree grows linearly in time, and the
growing node is chosen at each time from the growing number of active nodes.
In order to mimic the quenched noise of the medium$^{\text{\cite{h}}}$, one
can assign to each node of the tree an "effective" energy $\theta_{i}$
corresponding to the intermediate operator $\Theta$ by considering the strong
interaction from the environment, and require the higher energy nodes are more
likely to grow than lower energy ones.

In fact, the total Hamiltonian of the Cayley tree system plus environment now
is $H=H_{0}+\lambda H_{1}$, where $H_{0}=H_{S}+H_{B}$, $H_{S}$ is Hamiltonian
of the Cayley tree system, $H_{B}$ is Hamiltonian of the environment and
$H_{1}$ is supposed as an interaction part of the Hamiltonian $H$ by coupling
to the environment with coupling number $\lambda$ $>>1$. This kind of system
usually is difficult to treat using the perturbative method because the series
of expansion of the perturbation approach is related to the power of $\lambda
$, which is divergent. However, using the above proposed formula (\ref{28}), a
generalized reduced density operator for the Cayley tree system, $\rho_{proj}%
$, can be written as%
\begin{equation}
i\frac{\partial\rho_{proj}\left(  t\right)  }{\partial t}=\left[  \Theta
,\rho_{proj}\left(  t\right)  \right]  =\left[  H_{0}+\lambda H_{1}%
C,\rho_{proj}\left(  t\right)  \right]  , \label{5a}%
\end{equation}
which can give%
\begin{equation}
i\frac{\partial P}{\partial t}=P\Theta P=PH_{0}P+\lambda^{2}PH_{1}G_{Q}H_{1}P,
\label{5b}%
\end{equation}
with%
\begin{equation}
G_{Q}=\frac{1}{z-QHQ}. \label{5c}%
\end{equation}
By taking the Born expansion $G_{Q}=1+\lambda G_{Q}^{0}H_{1}+\lambda^{2}%
G_{Q}^{0}H_{1}G_{Q}^{0}H_{1}+\cdots$ one gets%
\begin{equation}
P\left(  \Theta-H_{0}\right)  P=\lambda PH_{1}\frac{1}{1-\lambda G_{Q}%
^{0}H_{1}}P. \label{5d}%
\end{equation}
When $\lambda>>1$, one obtain the corresponding eigenvalues of $\Theta$ as
\begin{equation}
\theta_{n}=\frac{1}{2}\left(  z_{n}^{0}+\left\langle \varphi_{n}\right\vert
H_{1}QH_{0}QH_{1}^{-1}\left\vert \varphi_{n}\right\rangle \right)  ,
\label{5e}%
\end{equation}
where $\varphi_{n}$ is the $n$th eigenvector of $PH_{0}P$, $H_{1}^{-1}$ is an
inverse operator of $H_{1}$, and noticing $\left\langle \varphi_{n}\right\vert
P\Theta P\left\vert \varphi_{n}\right\rangle =\theta_{n}$ is an eigenvalue for
the open system (the Cayley tree) by $P$ tracing out variables of the
environment. Therefore the probability $\rho_{i}$ for the active node $i$ with
energy $\theta_{i}$ to grow at time $t$ can be given by%
\begin{equation}
\rho_{i}=\frac{\exp\left(  -\frac{\beta_{proj}}{2}\left(  z_{i}^{0}%
+\left\langle \varphi_{i}\right\vert H_{1}QH_{0}QH_{1}^{-1}\left\vert
\varphi_{i}\right\rangle \right)  \right)  }{\sum_{j\in Int\left(  t\right)
}\exp\left(  -\frac{\beta_{proj}}{2}\left(  z_{j}^{0}+\left\langle \varphi
_{j}\right\vert H_{1}QH_{0}QH_{1}^{-1}\left\vert \varphi_{j}\right\rangle
\right)  \right)  }, \label{o14}%
\end{equation}
where the model depends on the parameter $\beta_{proj}$, which can change
characteristics of the tree. Comparing with the formula of $\rho_{i}$ in
original case$^{\text{\cite{h}}}$%
\begin{equation}
\rho_{i}=\frac{\exp\left(  -\beta z_{i}^{0}\right)  }{\sum_{j\in Int\left(
t\right)  }\exp\left(  -\beta z_{j}^{0}\right)  }, \label{o15}%
\end{equation}
it can be seen that the formula of this open system still has the similar
structure as the original one except the shift of phase $\left\langle
\varphi_{j}\right\vert H_{1}QH_{0}QH_{1}^{-1}\left\vert \varphi_{j}%
\right\rangle $ and $\beta\rightarrow\frac{\beta_{proj}}{2}$. This shows that
a node $i$ of the tree currently possesses an an "effective" energy
$\theta_{i}=z_{i}^{0}+\left\langle \varphi_{i}\right\vert H_{1}QH_{0}%
QH_{1}^{-1}\left\vert \varphi_{i}\right\rangle $, from the original energy
$z_{i}^{0}$, corresponding to a random distribution $p\left(  \theta\right)
$. This allows the Cayley tree network to remain, with similar characteristic
as that in the original case, if the interaction from environment can assign
to each node, such as node $i$, of the tree an "effective" energy $\theta
_{i}-z_{i}^{0}$ by keeping the original rule of forming Cayley tree (as a
resource of self-orgnization).

Again, consider a quantum network whose nodes are composite by (electron)
spins, $\sum_{\sigma}E_{d}n_{d}$, $\left(  n_{d\sigma}=d_{\sigma}^{+}%
d_{\sigma}\text{, }d_{\sigma}^{+}\text{ is creation operator of the
spin}\right)  $ with interactions as connections. The environment (or control)
field, $\sum_{k,\sigma}E_{k}n_{k\sigma}$, $\left(  n_{k\sigma}=C_{k\sigma}%
^{+}C_{k\sigma}\text{, }C_{k\sigma}^{+}\text{ is creation operator of the
fermi particle}\right)  $ are composed by infinite (electron) fermis. The
correlation between $n_{d}$ is $Un_{d\uparrow}n_{d\downarrow}=\frac{U}{2}%
\sum_{\sigma}n_{d\sigma}n_{d\sigma}$, where $U$ is correlation energy of
electrons. The interaction between the network and the environment is
$\lambda\sum_{k,\sigma}\left(  C_{k\sigma}^{+}d_{\sigma}+d_{\sigma}%
^{+}C_{k\sigma}\right)  $, where $\lambda$ is coupling number. Hence, the
Hamiltonian operator is expressed by%
\begin{equation}
H=\sum_{k,\sigma}E_{k}C_{k\sigma}^{+}C_{k\sigma}+\sum_{\sigma}E_{d}d_{\sigma
}^{+}d_{\sigma}+Un_{d\uparrow}n_{d\downarrow}+\lambda\sum_{k,\sigma}\left(
C_{k\sigma}^{+}d_{\sigma}+d_{\sigma}^{+}C_{k\sigma}\right)  . \label{t1}%
\end{equation}
When $\lambda=0$, the free Hamiltonian of the network gives two energy levels
as $E_{d}$ and $E_{d}+U$. This allows the nodes of the network to be possibly
in three combined status: $0$ ($E_{0}=0$), $1$ ($\sigma=\uparrow$ or
$\downarrow$, $E_{1\sigma}=E_{d}$), and $2$\ ($\sigma=\uparrow$ and
$\downarrow$, $2E_{d}+U$) occupations. Hence, following wolff
transformation$^{\text{\cite{23},\cite{24}}}$, we introduce three projectors
to divide the total Hilbert space as three subspaces:%
\begin{align}
P_{0}  &  =\left(  1-n_{d\uparrow}\right)  \left(  1-n_{d\downarrow}\right)
,\label{h1}\\
P_{1}  &  =n_{d\uparrow}\left(  1-n_{d\downarrow}\right)  +n_{d\downarrow
}\left(  1-n_{d\uparrow}\right)  ,\label{h2}\\
P_{2}  &  =n_{d\uparrow}n_{d\downarrow}, \label{h3}%
\end{align}
where $P_{n}$, $n=0,1,2$ corresponds upon $0$,$1$, $2$ occupations, with%
\begin{align}
P_{n}^{2}  &  =P_{n},\label{g1}\\
P_{n}P_{n^{\prime}}  &  =0. \label{g2}%
\end{align}
Suppose that the total wave function $\Psi$ is composed by $\psi_{0}$,
$\psi_{1}$, $\psi_{2}$ in the three subspaces, respectively, then the
Schr\"{o}dinger equation $H\Psi=E\Psi$ can be expressed as
\begin{equation}
\left(
\begin{array}
[c]{ccc}%
H_{00} & H_{01} & H_{02}\\
H_{10} & H_{11} & H_{12}\\
H_{20} & H_{21} & H_{22}%
\end{array}
\right)  \left(
\begin{array}
[c]{c}%
\psi_{0}\\
\psi_{1}\\
\psi_{2}%
\end{array}
\right)  =E\left(
\begin{array}
[c]{c}%
\psi_{0}\\
\psi_{1}\\
\psi_{2}%
\end{array}
\right)  \label{t2}%
\end{equation}
where $H_{nn^{\prime}}=P_{n}HP_{n^{\prime}}$, with
\begin{align}
H_{02}  &  =H_{20}=0,\label{j1}\\
H_{01}  &  =P_{0}HP_{1}=\lambda\sum_{k,\sigma}C_{k\sigma}^{+}\left(
1-n_{d\sigma}\right)  d_{\sigma},\label{j2}\\
H_{12}  &  =P_{1}HP_{2}=\lambda\sum_{k,\sigma}C_{k\sigma}^{+}d_{\sigma
}n_{d\sigma}. \label{j3}%
\end{align}
When the nodes are in the $1$ occupation, the local extra spins appear, which
allows the network shows type of local magnetic effect as a self organization
system. Thus cancelling $\psi_{0}$ and $\psi_{2}$ from Eq.(\ref{t2}) and
considering Eqs.(\ref{j1}) -(\ref{j3}), we obtain the SKE as%
\begin{equation}
\Theta_{1}\psi_{1}=E_{1}\psi_{1}, \label{j4}%
\end{equation}
where the intermediate operator $\Theta_{1}$ is given by
\begin{equation}
\Theta_{1}=H_{11}+H_{12}\left(  E-H_{22}\right)  ^{-1}H_{21}+H_{10}\left(
E-H_{00}\right)  ^{-1}H_{01}. \label{t3}%
\end{equation}
In the second approximation with respect to $\lambda$, we have
\begin{align}
H_{12}\left(  E-H_{22}\right)  ^{-1}H_{21}  &  =\sum_{k,k^{\prime}%
,\sigma,\sigma^{\prime}}\frac{\lambda^{2}}{U+E_{d}-E_{d^{\prime}}}\left(
-1\right)  C_{k\sigma}^{+}C_{k^{\prime}\sigma^{\prime}}d_{\sigma}n_{d\sigma
}d_{\sigma^{\prime}}^{+}n_{d\sigma^{\prime}},\label{t3b}\\
H_{10}\left(  E-H_{00}\right)  ^{-1}H_{01}  &  =\sum_{k,k^{\prime}%
,\sigma,\sigma^{\prime}}\frac{\lambda^{2}}{E_{d}-E_{k}}\left(  -1\right)
C_{k\sigma}^{+}C_{k^{\prime}\sigma^{\prime}}d_{\sigma^{\prime}}^{+}\left(
1-n_{d\sigma^{\prime}}\right)  d_{\sigma}\left(  1-n_{d\sigma^{\prime}%
}\right)  . \label{t3c}%
\end{align}
Considering $n_{d}=n_{d\uparrow}+n_{d\downarrow}=1,$ and $\widehat{S}%
^{z}=\frac{1}{2}\left(  d_{\uparrow}^{+}d_{\uparrow}-d_{\downarrow}%
^{+}d_{\downarrow}\right)  $, $\widehat{S}^{+}=d_{\uparrow}^{+}d_{\downarrow}%
$, $\widehat{S}^{-}=d_{\downarrow}^{+}d_{\uparrow}$, the Eqs.(\ref{t3b}) and
(\ref{t3c}) become to%
\begin{align}
H_{12}\left(  E-H_{22}\right)  ^{-1}H_{21}  &  =\sum_{k,k^{\prime}%
,\sigma,\sigma^{\prime}}\frac{\lambda^{2}}{U+E_{d}-E_{k^{\prime}}}\left\{
\left[  \widehat{S}^{z}\left(  C_{k\uparrow}^{+}C_{k^{\prime}\uparrow
}-C_{k\downarrow}^{+}C_{k^{\prime}\downarrow}\right)  +\right.  \right.
\label{t3d}\\
&  \left.  \left.  \widehat{S}^{z}C_{k\uparrow}^{+}C_{k^{\prime}\uparrow
}+\widehat{S}^{-}C_{k\uparrow}^{+}C_{k^{\prime}\downarrow}\right]  -\frac
{1}{2}\sum_{\sigma}C_{k\sigma}^{+}C_{k^{\prime}\sigma^{\prime}}\right\}
,\nonumber
\end{align}%
\begin{align}
H_{10}\left(  E-H_{00}\right)  ^{-1}H_{01}  &  =\sum_{k,k^{\prime}}%
\frac{\lambda^{2}}{E_{k}-E_{d}}\left\{  \left[  \widehat{S}^{z}\left(
C_{k\uparrow}^{+}C_{k^{\prime}\uparrow}-C_{k\downarrow}^{+}C_{k^{\prime
}\downarrow}\right)  +\right.  \right. \label{t3e}\\
&  \left.  \left.  \widehat{S}^{z}C_{k\uparrow}^{+}C_{k^{\prime}\uparrow
}+\widehat{S}^{-}C_{k\uparrow}^{+}C_{k^{\prime}\downarrow}\right]  +\frac
{1}{2}\sum_{\sigma}C_{k\sigma}^{+}C_{k^{\prime}\sigma^{\prime}}\right\}
.\nonumber
\end{align}
Replacing Eqs.(\ref{t3d}), (\ref{t3e}), and noticing $H_{11}\approx H_{0}%
=\sum_{k,\sigma}E_{k}C_{k\sigma}^{+}C_{k\sigma}$, Eq.(\ref{t3}) is given by%
\begin{equation}
\Theta_{1}=H_{0}+H_{p}+H_{ed} \label{tt0}%
\end{equation}
with%
\begin{equation}
H_{p}=\sum_{k,k^{\prime},\sigma}J_{kk^{\prime}}C_{k\sigma}^{+}C_{k^{\prime
}\sigma}\overset{U\rightarrow\infty}{\longrightarrow}J\sum_{k,k^{\prime
},\sigma}C_{k\sigma}^{+}C_{k^{\prime}\sigma} \label{v1}%
\end{equation}
and%
\begin{align}
H_{ed}  &  =-\sum_{k,k^{\prime}}J_{kk^{\prime}}\left\{  \widehat{S}^{z}\left(
C_{k\uparrow}^{+}C_{k^{\prime}\uparrow}-C_{k\downarrow}^{+}C_{k^{\prime
}\downarrow}\right)  +\widehat{S}^{z}C_{k\uparrow}^{+}C_{k^{\prime}\uparrow
}+\widehat{S}^{-}C_{k\uparrow}^{+}C_{k^{\prime}\downarrow}\right\}
\label{v2}\\
&  \overset{U\rightarrow\infty}{\longrightarrow}-J\sum_{k,k^{\prime}}\left\{
\widehat{S}^{z}\left(  C_{k\uparrow}^{+}C_{k^{\prime}\uparrow}-C_{k\downarrow
}^{+}C_{k^{\prime}\downarrow}\right)  +\widehat{S}^{z}C_{k\uparrow}%
^{+}C_{k^{\prime}\uparrow}+\widehat{S}^{-}C_{k\uparrow}^{+}C_{k^{\prime
}\downarrow}\right\}  , \label{v3}%
\end{align}
where in the strong correlation condition, $U\rightarrow\infty$, we have%
\begin{align}
J_{kk^{\prime}}  &  =\frac{\lambda^{2}}{2}\left\{  \frac{1}{E_{k}-E_{d}}%
-\frac{1}{U+E_{d}-E_{k^{\prime}}}\right\} \label{v4}\\
\overset{U\rightarrow\infty}{\longrightarrow}J  &  =-\frac{\lambda^{2}%
}{\left\vert E_{d}-E_{F}\right\vert }.\nonumber
\end{align}
On the other hand, by means of the double time delay Green function (following
ref.\cite{j})%
\begin{equation}
\left\langle \left\langle A\left(  t\right)  ;B\left(  t^{\prime}\right)
\right\rangle \right\rangle =-iu\left(  t-t^{\prime}\right)  \left\langle
\left[  A\left(  t\right)  ,B\left(  t^{\prime}\right)  \right]
_{+}\right\rangle , \label{v5}%
\end{equation}
and the Furiour transformation%
\begin{equation}
\left\langle \left\langle A\right.  \right.  \left\vert \left.  B\right\rangle
\right\rangle _{\omega}=\int dte^{i\omega\left(  t-t^{\prime}\right)
}\left\langle \left\langle A\left(  t\right)  ;B\left(  t^{\prime}\right)
\right\rangle \right\rangle \label{v8}%
\end{equation}
one have%
\begin{align}
\omega\left\langle \left\langle A\right.  \right.  \left\vert \left.
B\right\rangle \right\rangle _{\omega}  &  =\left\langle \left[  A,B\right]
_{+}\right\rangle +\left\langle \left\langle \left[  A,H\right]  \right.
\right.  \left\vert \left.  B\right\rangle \right\rangle _{\omega}%
,\label{v9}\\
\omega\left\langle \left\langle A\right.  \right.  \left\vert \left.
B\right\rangle \right\rangle _{\omega}  &  =\left\langle \left[  A,B\right]
_{+}\right\rangle -\left\langle \left\langle \left[  A,H\right]  \right.
\right.  \left\vert \left.  B\right\rangle \right\rangle _{\omega}.
\label{v10}%
\end{align}
All these allows one to solve the Eq.(\ref{tt0}) to obtain
\begin{align}
\left\langle n_{d\uparrow}\right\rangle  &  =\frac{1}{\pi}%
\operatorname{arccot}\left[  \frac{E_{d}-E_{F}+U\left\langle n_{d\downarrow
}\right\rangle }{\Gamma}\right]  ,\label{u1}\\
\left\langle n_{d\downarrow}\right\rangle  &  =\frac{1}{\pi}%
\operatorname{arccot}\left[  \frac{E_{d}-E_{F}+U\left\langle n_{d\downarrow
}\right\rangle }{\Gamma}\right]  , \label{u2}%
\end{align}
where $\Gamma=V^{2}\rho^{\left(  0\right)  }\left(  E_{F}\right)  $ represents
half width, in which the energy of the system distribute around the resonance
state $E_{d}+U\left\langle n_{d\downarrow}\right\rangle $ with width $2\Gamma
$, and the eigenvalue of the $\Theta$ is solved as a complex number
\[
\theta_{d\sigma}=E_{d}+U\left\langle n_{d\downarrow}\right\rangle +i\Gamma.
\]
When $E_{d}<E_{F}$, $E_{d}+U>E_{F}$, and $\left\vert E_{d}+U-E_{F}\right\vert
$, $\left\vert E_{d}-E_{F}\right\vert >>\Gamma$, there exist a local solution
for the Eqs.(\ref{u1}) and (\ref{u2}): $\left\langle n_{d\uparrow
}\right\rangle \rightarrow1$, $\left\langle n_{d\downarrow}\right\rangle
\rightarrow0$, the network appears to have local magnetic vectors as a type of
self-organization structure by interaction with huge nose environment. Its
quantum statistical distribution in the (generalized) canonical ensemble can
be given by previous formalsim as
\begin{equation}
\rho\left(  \theta_{d\sigma}\right)  =Z^{-1}e^{-\beta_{proj}\theta_{d\sigma}},
\label{t8}%
\end{equation}
and%
\begin{equation}
Z=\sum_{d,\sigma}\exp\left(  -\beta_{proj}\theta_{d\sigma}\right)  .
\label{t9}%
\end{equation}
The irreversible generalized force and the entropy for the network system can
be given by the above canonical partition function $Z$ formulation, i.e.%
\begin{equation}
Y=-\frac{1}{\beta}\frac{\partial}{\partial y}\ln Z, \label{o1}%
\end{equation}
and%
\begin{equation}
S=k\left(  \ln Z-\beta_{proj}\frac{\partial}{\partial\beta_{proj}}\ln
Z\right)  , \label{o2}%
\end{equation}
which shows immediately that $Y$ and $S$ are complex since $Z$ including
complex eigenvalue $\theta_{d\sigma}$. This means that the entropy of this
irreversible system is complex! Usually, the extension of the Hilbert space
technique can be derived the complex spectrum for the self-adjoint operator,
which demonstrates that the evolution of the intrinsic irreversible system has
two semigroups to represent asymmetric time evolution, which have been
discussed by many publications$^{\text{\cite{r1}}}$. What is new here, through
constructing the non-equilibrium ensembles formulas, we simply reveal that the
complex spectrum of the $\Theta$ can introduce complex entropy and complex
generalized force, which should be a characteristic of irreversible system,
and hard to find by using the equilibrium ensemble formalism of Gibbs.

\section{Conclusions}

In conclusions, we have proposed general non-equilibrium ensembles formalism
based on the subdynamic equation. The constructed procedure is to use a
similarity transformation between Gibbsian ensembles formalism based on the
Liouville equation and the non-equilibrium ensemble formalism based on SKE.
The obtained density distribution formula is a projected one that can
represent essence part of (irreversible) evolution of the density
distribution. Using this formulation, the irreversibility of the
non-equilibrium system may emerge naturally as its entropy becomes complex,
and can be exposed by calculating its general reduced density distribution.

\end{document}